%% file: rl_1.tex
\documentclass[11pt]{article}
\usepackage{moriond,epsfig}

\begin{document}
\vspace*{4cm}
\title{CORRELATED MIXTURES OF ADIABATIC AND ISOCURVATURE COSMOLOGICAL
PERTURBATIONS}

\author{A. RIAZUELO, D. LANGLOIS}

\address{D\'epartement d'astrophysique relativiste et de cosmologie,
UMR 8629 du CNRS, \\ Observatoire de Paris, F-92195 Meudon, France}

\maketitle\abstracts{
The Cosmic Microwave Background (CMB) anisotropies measurements can
provide many clues about the Universe. Although the common belief is
that they will allow a very precise measurement of the cosmological
parameters (that is, the current state of the Universe), they will
alternatively give interesting informations about the state of the
initial perturbations (that is, the state of the Universe at the end
of inflation). In this paper, we study the observational
consequences on the CMB anisotropies of some wide set of inital
conditions, with a correlated mixture of adiabatic and isocurvature
perturbations.}

\section{Introduction}

The CMB anisotropies can indirectly measure the cosmological
parameters by looking at the evolution of the cosmological
perturbations between the end of inflation and the recombination
epoch. In order to do so, one must implicitly assume a simple form for
the initial perturbations. Usually, one considers only adiabatic
fluctuations, with a power law spectrum.

Adiabatic fluctuations arise in the context of the simplest
inflationary scenario. However, as soon as one has a multiple
inflation scenario, isocurvature fluctuations can be generated, the
amplitude of which, as well as their correlation with the adiabatic
part, depend on the parameters of the model. In many of the models
already studied~\cite{iso}, the isocurvature and the adiabatic parts
of the fluctuations are uncorrelated, but it is possible to have a
correlated mixture of such perturbations, as was already stressed by
one of us~\cite{l99} in the study of a specific inflation model with
two massive non interacting scalar fields.

In this communication, we study this issue in a phenomenological way,
and look at the observable consequences of a correlated mixture of
adiabatic and isocurvature cosmological perturbations. These
perturbations happen to have a richer structure than the more usual
uncorrelated adiabatic and isocurvature ones.

\section{Definitions and notations}

When considering a mixture of several fluids, one can define an
entropy perturbation $S_{A,B}$ for any pair of components $A$ and $B$.
It is non zero as soon as the different particle number density
contrasts $\delta n_X/n_X$ ($X=A,B$), are not equal. It can also be
written in terms of energy density contrasts $\delta_X$~:
\begin{equation}
S_{A,B} \equiv \frac{\delta n_A}{n_A} - \frac{\delta n_B}{n_B}
 = \frac{\delta_A}{1+\omega_A} - \frac{\delta_B}{1+\omega_B},
\end{equation}
where $\omega_X \equiv p_X / \rho_X$ is the equation of state
parameter for the species $X$. (Note that this definition is
gauge-invariant.) Adiabatic initial conditions are defined such that
all the entropy perturbations are zero.  In cosmology, the
perturbations are actually considered as random fields, usually
assumed to be Gaussian, and described by their power spectra. When one
has to deal with several random fields, one must also impose the form
of the cross-correlation between them. In what follows, we will
consider only \emph{totally correlated} adiabatic and isocurvature
perturbations, where all the random fields are described in terms of a
single random variable. We further assume that only one species
deviates from adiabaticity. In this case, one simply has to define
which species deviates from adiabaticity, say $X$ (i.e.~$S_{A,B} = 0$
when $A,B\neq X$), and the relative initial amplitude between the
entropy perturbation $S_{X,Y}$ and the Bardeen potential $\Phi$ (where
$Y$ is another species which does not deviate from adiabaticity)~:
\begin{equation}
S_{X,Y} \equiv \lambda \Phi.
\end{equation}
We will then talk about ``$X$ hybrid perturbation''.

\section{An analytical estimate}

Adiabatic scale invariant initial conditions make two predictions
concerning the CMB anisotropies. First, they predict a flat
(Sachs-Wolfe) plateau at low multipoles, which illustrates the fact
that the gravitational potential is ``frozen'' as long as the modes
have not yet entered into the Hubble radius. Second, one expects to
find a serie of Doppler peaks at smaller angular scales, produced by
acoustic oscillations in the photon-baryon plasma. The height of the
first peak depends on almost all the cosmological parameters, but as
soon as one considers adiabatic initial conditions and (very)
conservative cosmological parameters, the peak is between $2$ and $8$
times higher than the Sachs-Wolfe plateau. This is a strong prediction
of adiabatic models, and it is in good agreement with the current
data. (In opposition, a pure isocurvature CDM model leads to a
Sachs-Wolfe plateau higher than the first Doppler peak.) In the case
of adiabatic and isocurvature mixtures, the CMB anisotropies and the
matter power spectrum correspond, on large scales, to different
combinations of the initial perturbations.  Such a complementarity is
all the more useful in our model that, contrarily to the adiabatic
case, it does not generically predict the ratio of amplitude between
the CMB and the matter power spectrum. Indeed, in the standard
adiabatic case, it is a classic calculation to derive the temperature
anisotropy as a function of the gravitational potential. In the long
wavelength limit, the temperature anisotropies are one third of the
gravitational potential, which has varied of a factor $\simeq 9/10$
during the radiation-to-matter transition. In the case of CDM hybrid
perturbations, this relation can be rewritten as~:
\begin{equation}
\left.\frac{\delta T}{T}\right|_{\mathrm {MD}}
 = \frac{3}{10}
 \left(1+\frac{4}{15}\Omega_\nu^{\mathrm{RD}}
 - \frac{2}{5}\lambda \Omega_c^{\mathrm{MD}}\right)
 \Phi_{\mathrm {RD}},
\end{equation}
where $\Omega_c$ and $\Omega_\nu$ are respectively the CDM and
neutrino density parameters, and the indexes $\mathrm {RD}$ and
$\mathrm {MD}$ mean that one considers the quantities during the
radiation dominated and the matter dominated eras respectively. The
relative amplitude of CMB anisotropies and matter power spectrum can
therefore in principle be used to extract the isocurvature part of the
initial conditions.

\section{Numerical results}

We have extensively studied the four hybrid perturbations in a recent
paper~\cite{lr1999}. The main result of our analysis is that the CMB
anisotropy and the matter power spectra and rather strongly sensitive
to the set of initial conditions we have considered. As a consequence,
such models are already fairly well constrained~: one cannot deviate
strongly from adiabaticity. As an example, we have plotted on
Fig.~\ref{fig_pic} the relative height of the first Doppler peak in
the four hybrid models one can consider. Photon and CDM hybrid
perturbations are the most constrained. As for the relative amplitude
between the CMB anisotropies and the matter power spectrum, we have
compared them on Fig.~\ref{fig_dtps}. It is clear that the amplitude
of the first Doppler peak, which is already well measured by several
ground and balloon experiments, is strongly sensitive to the parameter
$\lambda$. This might help to discriminate between these models and
the standard adiabatic model.

\begin{figure}
\centering
\input{pic1_plat}
\caption{Ratio of the height of the first Doppler peak to the
Sachs-Wolfe plateau in the four hybrid models. We have introduced the
variable $\theta$, defined as $\delta_X \equiv \delta_X^{\rm adia}
\cot \theta$, where $\delta_X^{\rm adia}$ is the value of the density
contrast of the species $X$ which deviates from adiabaticity in the
corresponding adiabatic model. The other cosmological parameters are
those of the SCDM model.}
\label{fig_pic}
\end{figure}
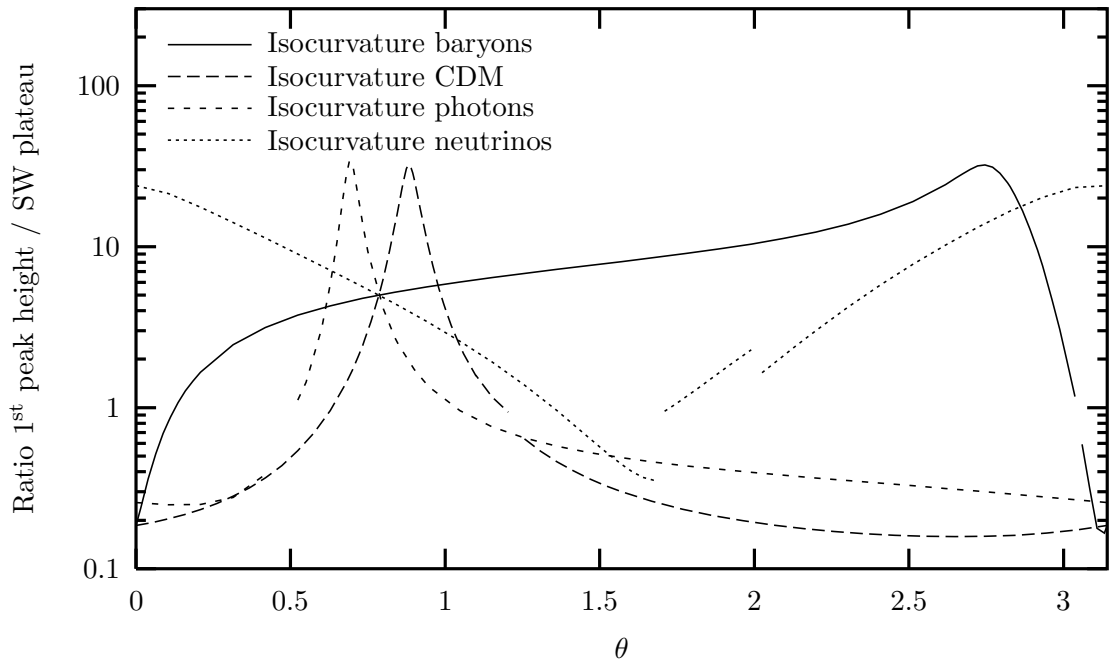
\begin{figure}
\centering
\input{cor1dt}
\input{cor1ps}
\caption{Comparison of the CMB anisotropy (top) and matter power
spectra (bottom) in various hybrid CDM models. All the models have
been COBE-normalised. Note the difference between the Doppler peak
height relatively with the Sachs-Wolfe plateau, as well as the
different normalisation of the matter power spectrum. The position of
the Doppler peaks as well as the position of the maximum in the matter
power spectrum also slightly vary with the parameters. }
\label{fig_dtps}
\end{figure}
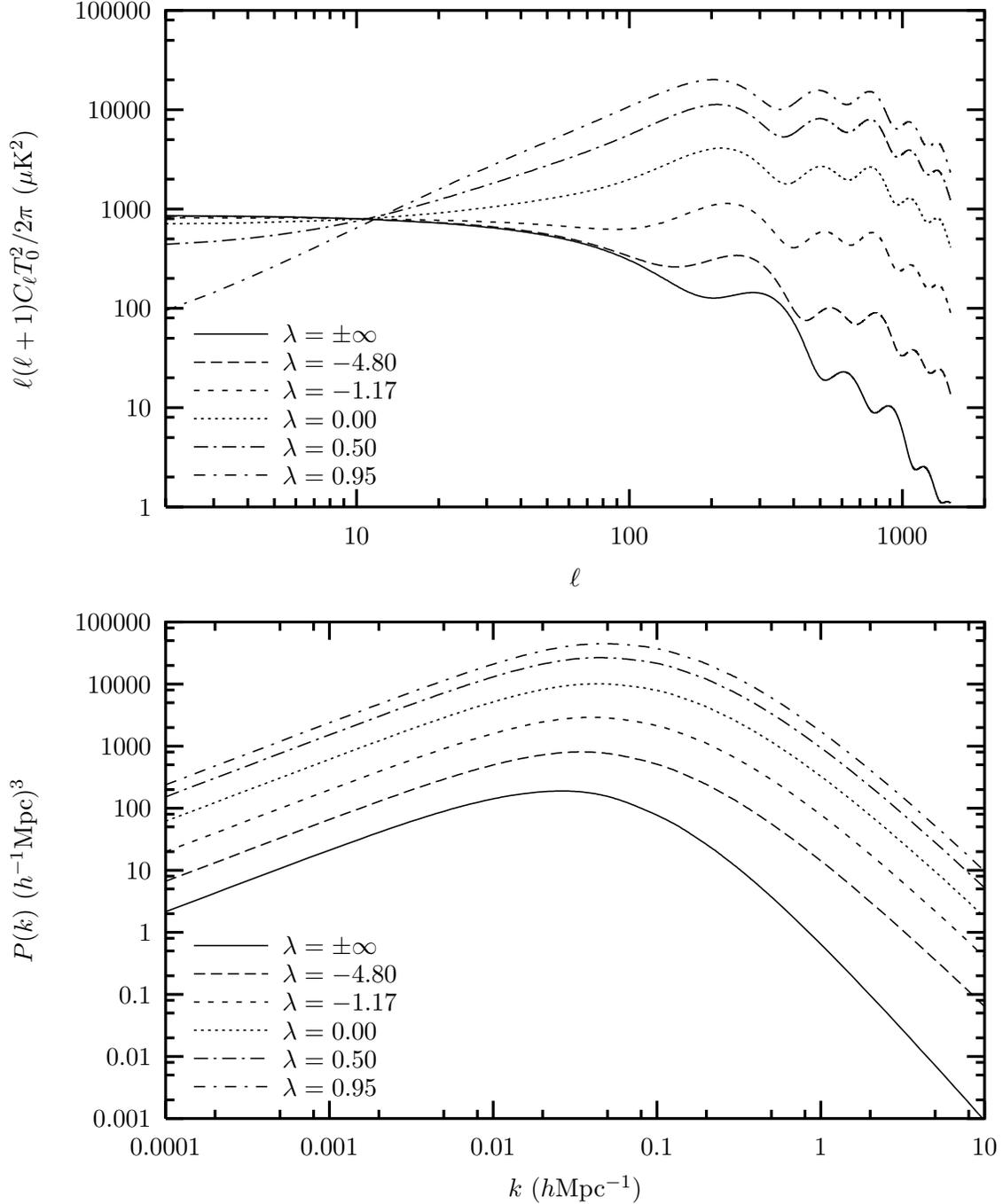

\section*{References}
\newcommand{\JOURNAL}[4]{{\it{#1}} {\bf{#2}}, {#3} ({#4})}

\end{document}

%% file: pic1_plat.tex
% GNUPLOT: LaTeX picture with Postscript
\begingroup%
  \makeatletter%
  \newcommand{\GNUPLOTspecial}{%
    \@sanitize\catcode`\%=14\relax\special}%
  \setlength{\unitlength}{0.12bp}%
\begin{picture}(3600,2160)(0,0)%
\special{psfile=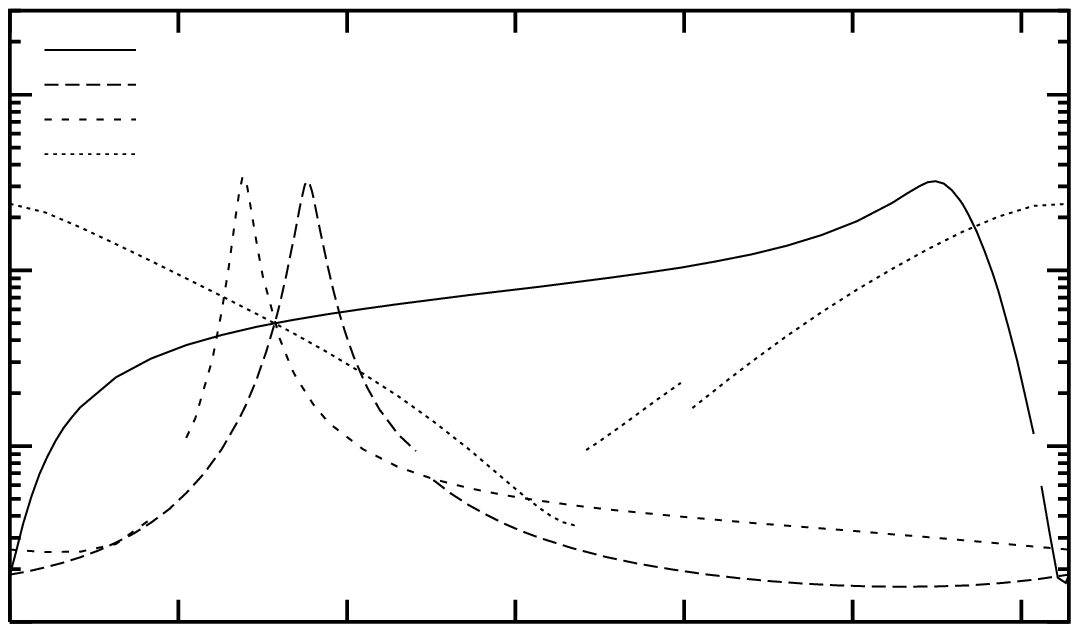 llx=0 lly=0 urx=600 ury=420 rwi=7200}
\put(813,1647){\makebox(0,0)[l]{Isocurvature neutrinos}}%
\put(813,1747){\makebox(0,0)[l]{Isocurvature photons}}%
\put(813,1847){\makebox(0,0)[l]{Isocurvature CDM}}%
\put(813,1947){\makebox(0,0)[l]{Isocurvature baryons}}%
\put(1925,50){\makebox(0,0){ $\theta$ }}%
\put(100,1180){%
\special{ps: gsave currentpoint currentpoint translate
270 rotate neg exch neg exch translate}%
\makebox(0,0)[b]{\shortstack{Ratio 1$^{\rm st}$ peak height / SW plateau}}%
\special{ps: currentpoint grestore moveto}%
}%
\put(3313,200){\makebox(0,0){3}}%
\put(2827,200){\makebox(0,0){2.5}}%
\put(2342,200){\makebox(0,0){2}}%
\put(1856,200){\makebox(0,0){1.5}}%
\put(1371,200){\makebox(0,0){1}}%
\put(885,200){\makebox(0,0){0.5}}%
\put(400,200){\makebox(0,0){0}}%
\put(350,1818){\makebox(0,0)[r]{100}}%
\put(350,1312){\makebox(0,0)[r]{10}}%
\put(350,806){\makebox(0,0)[r]{1}}%
\put(350,300){\makebox(0,0)[r]{0.1}}%
\end{picture}%
\endgroup
 

%% file: cor1dt.tex
% GNUPLOT: LaTeX picture with Postscript
\begingroup%
  \makeatletter%
  \newcommand{\GNUPLOTspecial}{%
    \@sanitize\catcode`\%=14\relax\special}%
  \setlength{\unitlength}{0.12bp}%
\begin{picture}(3600,2160)(0,0)%
\special{psfile=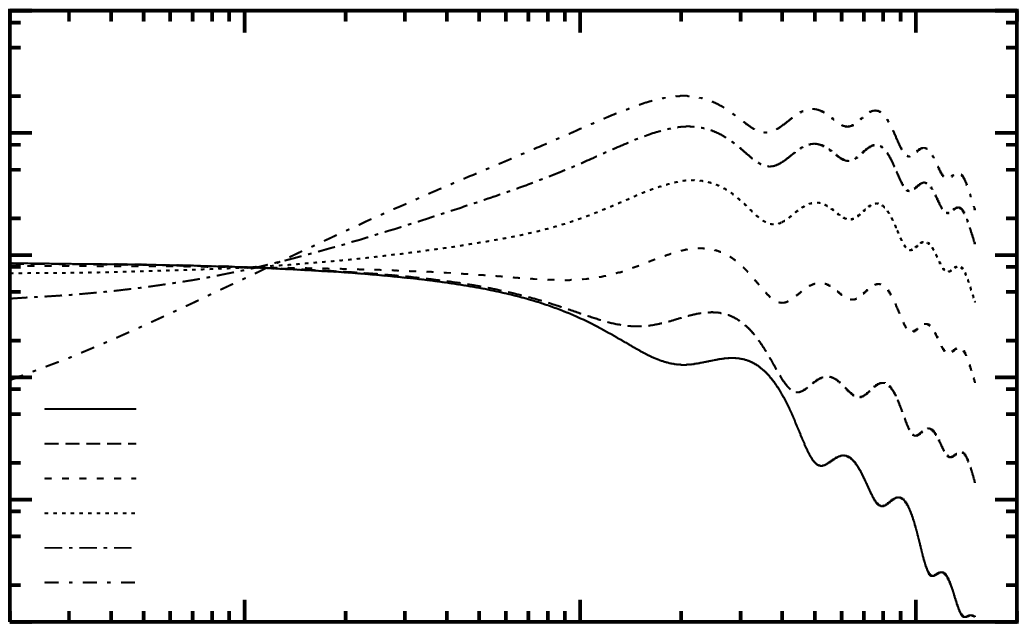 llx=0 lly=0 urx=600 ury=420 rwi=7200}
\put(963,413){\makebox(0,0)[l]{$\lambda = 0.95$}}%
\put(963,513){\makebox(0,0)[l]{$\lambda = 0.50$}}%
\put(963,613){\makebox(0,0)[l]{$\lambda = 0.00$}}%
\put(963,713){\makebox(0,0)[l]{$\lambda = -1.17$}}%
\put(963,813){\makebox(0,0)[l]{$\lambda = -4.80$}}%
\put(963,913){\makebox(0,0)[l]{$\lambda = \pm\infty$}}%
\put(2000,50){\makebox(0,0){$\ell$}}%
\put(100,1180){%
\special{ps: gsave currentpoint currentpoint translate
270 rotate neg exch neg exch translate}%
\makebox(0,0)[b]{\shortstack{$\ell(\ell+1) C_\ell T_0^2 / 2 \pi$ ($\mu$K$^2$)}}%
\special{ps: currentpoint grestore moveto}%
}%
\put(3159,200){\makebox(0,0){1000}}%
\put(2192,200){\makebox(0,0){100}}%
\put(1226,200){\makebox(0,0){10}}%
\put(500,2060){\makebox(0,0)[r]{100000}}%
\put(500,1708){\makebox(0,0)[r]{10000}}%
\put(500,1356){\makebox(0,0)[r]{1000}}%
\put(500,1004){\makebox(0,0)[r]{100}}%
\put(500,652){\makebox(0,0)[r]{10}}%
\put(500,300){\makebox(0,0)[r]{1}}%
\end{picture}%
\endgroup
 

%% file: cor1ps.tex
% GNUPLOT: LaTeX picture with Postscript
\begingroup%
  \makeatletter%
  \newcommand{\GNUPLOTspecial}{%
    \@sanitize\catcode`\%=14\relax\special}%
  \setlength{\unitlength}{0.12bp}%
\begin{picture}(3600,2160)(0,0)%
\special{psfile=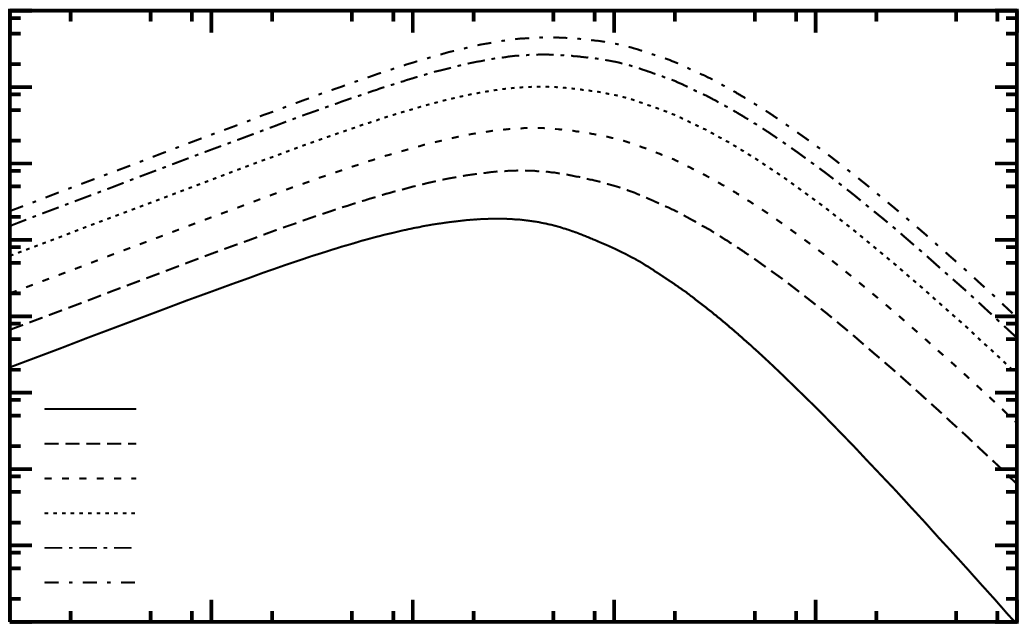 llx=0 lly=0 urx=600 ury=420 rwi=7200}
\put(963,413){\makebox(0,0)[l]{$\lambda = 0.95$}}%
\put(963,513){\makebox(0,0)[l]{$\lambda = 0.50$}}%
\put(963,613){\makebox(0,0)[l]{$\lambda = 0.00$}}%
\put(963,713){\makebox(0,0)[l]{$\lambda = -1.17$}}%
\put(963,813){\makebox(0,0)[l]{$\lambda = -4.80$}}%
\put(963,913){\makebox(0,0)[l]{$\lambda = \pm\infty$}}%
\put(2000,50){\makebox(0,0){$k$ ($h$Mpc$^{-1}$)}}%
\put(100,1180){%
\special{ps: gsave currentpoint currentpoint translate
270 rotate neg exch neg exch translate}%
\makebox(0,0)[b]{\shortstack{$P(k)$ ($h^{-1}$Mpc)$^3$}}%
\special{ps: currentpoint grestore moveto}%
}%
\put(3450,200){\makebox(0,0){10}}%
\put(2870,200){\makebox(0,0){1}}%
\put(2290,200){\makebox(0,0){0.1}}%
\put(1710,200){\makebox(0,0){0.01}}%
\put(1130,200){\makebox(0,0){0.001}}%
\put(550,200){\makebox(0,0){0.0001}}%
\put(500,2060){\makebox(0,0)[r]{100000}}%
\put(500,1840){\makebox(0,0)[r]{10000}}%
\put(500,1620){\makebox(0,0)[r]{1000}}%
\put(500,1400){\makebox(0,0)[r]{100}}%
\put(500,1180){\makebox(0,0)[r]{10}}%
\put(500,960){\makebox(0,0)[r]{1}}%
\put(500,740){\makebox(0,0)[r]{0.1}}%
\put(500,520){\makebox(0,0)[r]{0.01}}%
\put(500,300){\makebox(0,0)[r]{0.001}}%
\end{picture}%
\endgroup
 